\documentclass[twocolumn]{aastex63}

\newcommand{\fesc}{\ifmmode{f_{\rm esc}}\else{$f_{\rm esc}$}\fi}
\newcommand{\fescs}{\ifmmode{f_{\rm esc}^\star}\else{$f_{\rm esc}^\star$}\fi}
\newcommand{\kms}{\ifmmode{{\;\rm km~s^{-1}}}\else{km~s$^{-1}$}\fi}
\newcommand{\fgas}{\ifmmode{{f_{\rm gas}}}\else{$f_{\rm gas}$}\fi}
\newcommand{\cubecm}{\ifmmode{{\rm cm^{-3}}}\else{cm$^{-3}$}\fi}
\newcommand{\ztwo}{\ifmmode{{\rm [Z_2/H]}}\else{[Z$_2$/H]}\fi}
\newcommand{\zthree}{\ifmmode{{\rm [Z_3/H]}}\else{[Z$_3$/H]}\fi}
\newcommand{\lsim}{\lower0.3em\hbox{$\,\buildrel <\over\sim\,$}}
\newcommand{\gsim}{\lower0.3em\hbox{$\,\buildrel >\over\sim\,$}}

\newcommand{\sfr}{\ifmmode{\textrm{M}_\odot \,\textrm{yr}^{-1} \,\textrm{Mpc}^{-3}}\else{M$_\odot$ yr$^{-1}$ Mpc$^{-3}$}\fi}
\newcommand{\hsfr}{\ifmmode{\textrm{M}_\odot\, \textrm{yr}^{-1}}\else{M$_\odot$ yr$^{-1}$}\fi}

\newcommand{\eavg}{\ifmmode{\langle E_\gamma \rangle}\else{$\langle E_\gamma \rangle$}\fi}

\newcommand{\Ms}{\ifmmode{M_\odot}\else{$M_\odot$}\fi}
\newcommand{\vrms}{\ifmmode{v_{\rm rms}}\else{$v_{\rm rms}$}\fi}

\newcommand{\tvir}{\ifmmode{T_{\rm{vir}}}\else{$T_{\rm{vir}}$}\fi}
\newcommand{\mvir}{\ifmmode{M_{\rm{vir}}}\else{$M_{\rm{vir}}$}\fi}
\newcommand{\rvir}{\ifmmode{r_{\rm{vir}}}\else{$r_{\rm{vir}}$}\fi}

\newcommand{\jj}{\ifmmode{J_{21}}\else{$J_{21}$}\fi}
\newcommand{\flw}{\ifmmode{F_{LW}}\else{$F_{LW}$}\fi}
\newcommand{\kph}{\ifmmode{k_{\rm ph}}\else{$k_{\rm ph}$}\fi}

\newcommand{\zsun}{\ifmmode{\rm\,Z_\odot}\else{$\rm\,Z_\odot$}\fi}

\newcommand{\hi}{H {\sc i}}
\newcommand{\hii}{H {\sc ii}}
\newcommand{\hei}{He {\sc i}}
\newcommand{\heii}{He {\sc ii}}
\newcommand{\heiii}{He {\sc iii}}
\newcommand{\nhi}{\ifmmode{N_{\rm HI}}\else{$N_{\rm HI}$}\fi}

\newcommand{\oiii}{[O {\sc iii}]}
\newcommand{\oii}{[O {\sc ii}]}

\submitjournal{ApJL}
\begin{document}

\title{The Lyman Continuum Escape Survey: Connecting Time-Dependent \oiii\ and \oii\ Line Emission with Lyman Continuum Escape Fraction in Simulations of Galaxy Formation}

\correspondingauthor{Kirk S. S. Barrow}
\email{kssbarrow@gmail.com}

\author[0000-0002-8638-1697]{Kirk S. S. Barrow}
\affiliation{Kavli Institute for Particle Astrophysics and Cosmology,  Stanford University, 452 Lomita Mall, Stanford, CA  94305-4085, USA}

\author[0000-0002-4271-0364]{Brant E. Robertson}
\affiliation{Department of Astronomy \& Astrophysics, University of California, Santa Cruz, 1156 High St., Santa Cruz, CA 95064, USA}

\author[0000-0001-7782-7071]{Richard S. Ellis}
\affiliation{Department of Physics and Astronomy, University College London, Gower Street, London WC1E 6BT, UK}

\author[0000-0003-2965-5070]{Kimihiko Nakajima}
\affiliation{National Astronomical Observatory of Japan, 2-21-1 Osawa, Mitaka, Tokyo 181-8588, Japan}

\author[0000-0001-5333-9970]{Aayush Saxena}
\affiliation{Department of Physics and Astronomy, University College London, Gower Street, London WC1E 6BT, UK}

\author{Daniel P. Stark}
\affiliation{Steward Observatory, University of Arizona, 933 N. Cherry Ave., Tucson, AZ 85721, USA}

\author{Mengtao Tang}
\affiliation{Steward Observatory, University of Arizona, 933 N. Cherry Ave., Tucson, AZ 85721, USA}
\affiliation{Department of Physics and Astronomy, University College London, Gower Street, London WC1E 6BT, UK}

 \shortauthors{Barrow et al.}
 \shorttitle{Lyman Continuum Escape and Optical Line Emission}

\begin{abstract}

Escaping Lyman continuum photons from galaxies likely reionized the intergalactic medium at redshifts $z\gtrsim6$. However, the Lyman continuum is not directly observable at these redshifts and secondary indicators of Lyman
continuum escape must be used to estimate the budget of ionizing photons. Observationally, at redshifts
$z\sim2-3$ where the Lyman continuum is observationally accessible, surveys have established that many objects
that show appreciable Lyman continuum escape fractions $f_{esc}$ also show enhanced \oiii$\,/\,$\oii\ (O$_{32}$) emission line ratios.
Here, we use radiative transfer analyses of cosmological zoom-in simulations of galaxy formation to study the physical
connection between $f_{esc}$ and O$_{32}$. Like the observations, we find that the largest $f_{esc}$ values
occur at elevated O$_{32}\sim3-10$ and that the combination of high $f_{esc}$ and low O$_{32}$ is extremely
rare. While high $f_{esc}$ and O$_{32}$ often are observable concurrently, the timescales of the physical origin
for the
processes are very different. Large O$_{32}$ values fluctuate on short ($\sim$1 Myr) timescales during
the Wolf-Rayet-powered phase after the formation of star clusters,
while channels of low absorption are established over tens of megayears by collections of supernovae.
We find that while there is no direct causal relation between $f_{esc}$ and O$_{32}$, high $f_{esc}$
most often occurs after continuous input from star formation-related feedback events that have
corresponding excursions to large O$_{32}$ emission. These calculations are in agreement with
interpretations of observations that large $f_{esc}$ tends to occur when O$_{32}$ is large, but
large O$_{32}$ does not necessarily imply efficient Lyman continuum escape.

\end{abstract}

%\begin{multicols}{2}

\section{Introduction}

The process of cosmic reionization represents a major challenge for understanding
the large-scale evolution of the intergalactic medium (IGM). Reionization completed
during the first billion years of cosmic history, as evidenced by the prominent
\citet{gunn1965a} absorption troughs from neutral hydrogen observed in the spectra of
quasars at redshifts $z>6$ \citep{fan2001a,fan2006a,banados2018a}. Given the
rapid decline in the abundance of bright quasars over the same epoch, star forming
galaxies at high redshift likely produced the Lyman continuum photons required
to reionize the IGM \citep[][]{robertson2015a,bouwens2015a,finkelstein2019a}.
The opacity of the mostly ionized IGM at late times remains high enough to
prevent the direct detection of Lyman continuum (LyC) photons much beyond redshift $z\sim3$
\citep[][]{madau1995a,steidel2001a,inoue2014a}.
During the reionization epoch,
probes of the potential LyC
production and escape must rely on secondary observational indicators, such
as nebular emission lines from
galaxies, that the forthcoming \emph{James Webb
Space Telescope} (\emph{JWST}) will examine in detail.
Motivated by the need to understand the physics behind secondary
indicators of LyC escape, this {\emph Letter} presents
radiative transfer calculations
in high-resolution hydrodynamical simulations of galaxy formation to study the
connection between LyC escape fraction $f_{\rm{esc}}$ and rest-frame optical emission
lines powered by ionizing radiation from massive stars.

Given the importance of understanding how galaxies
might reionize the IGM, the search for evidence of escaping LyC photons has been
wide-ranging. Searches of nearby galaxies have detected
LyC emission in some unsually compact or star-bursting galaxies
\citep{borthakur2014a,izotov2016a,izotov2016b,leitherer2016a}.
Blue-sensitive spectrographs have provided direct spectroscopic
evidence for LyC emission \citep{steidel2001a,shapley2006a,steidel2018a},
as has ground-based continuum imaging
\citep{iwata2009a,vanzella2010a,nestor2011a,nestor2013a,mostardi2013a,grazian2016a,mestric2020a}.
Owing to the need for high-resolution imaging in identifying potential foreground
contamination \citep{vanzella2012a,mostardi2015a},
many recent searches for LyC
have focused on redshifts $z\sim2-3$ where ultraviolet (UV) filters on \emph{Hubble
Space Telescope} (\emph{HST})
probe blueward of $912\AA$ in the galaxy rest frame. These efforts include our
LymAn Continuum Escape Survey \citep[LACES, HST GO-14747;][]{fletcher2019a}
that has to date focused on observational connections between LyC escape and
the ionizing photon production evidenced by nebular line emission in galaxies
\citep{nakajima2020a}. These direct searches have been complemented by studies
of the association of LyC production with ultraviolet or optical
nebular lines \citep{tang2019a,du2020a}, the correspondence between
Lyman-$\alpha$ and the
(\oiii$\lambda$5007 + \oiii$\lambda$4959 )/\oii$\lambda$3727 line ratio (O$_{32}$) \citep{izotov2020a}, and the link between LyC
escape, H$\beta$ emission, and the rest-frame UV spectral slope \citep{yamanaka2020a}.

Relating LyC and optical emission lines at high redshift currently requires
infrared spectrographs on ground-based large telescopes that can access
redshifted rest-frame ultraviolet and
optical lines \citep{nakajima2016a,nakajima2018a}.
Studies of the LyC-line emission connection are
motivated in part by the analysis by \citet{2013ApJ...766...91J,nakajima2014a},
who suggested that the structure of photoionization regions within a
galaxy may induce a connection between $f_{\rm{esc}}$ and O$_{32}$. Many galaxies with LyC detections at redshifts
$z\sim2-3$ do indeed show elevated O$_{32}$ and combined (\oiii$\lambda$5007 + \oiii$\lambda$4959 + \oii$\lambda$3727)/H$\beta$ $\lambda$4861 measure
known as R$_{23}$, but not all strong line emitters display escaping LyC
\citep{naidu2018a,jaskot2019a,bassett2019a} and active galactic nuclei may
contribute to a portion that do \citep{smith2018a,smith2020a}.

In this Letter, radiative transfer calculations are applied to
simulations of galaxy formation to study how LyC escape and optical
line emissions are physically connected.
The radiative transfer of LyC photons from galaxies has
been examined in cosmological simulations of galaxy formation
\citep[e.g.,][]{ma2016a,trebitsch2017a},
where feedback from star formation was shown to play an important
role in enabling hydrogen ionizing photons to escape into the IGM.
Previous studies are extended by additionally
examining time-dependent \oiii and \oii line emission, and their
relation to the ionizing photon production of newly-formed stars
\citep[see also][]{katz2020a}. We show for the first time that galaxy population statistics of  $f_{\rm{esc}}$ and O$_{32}$ may be explained by these time-dependent processes.

\section{Methods}
\label{section:methods}

\subsection{Cosmological Simulation}

Results are based on a radiation-hydrodynamic adaptive mesh refinement {\sc Enzo}  \citep{2014ApJS..211...19B} simulation evolved from initial conditions produced as part of the {\sc Agora} collaboration \citep{2014ApJS..210...14K}. The simulation is run with cosmological parameters $\Omega_M = 0.3065$, $\Omega_{\Lambda} = 0.6935 $, $\Omega_b = 0.0483$, $h = 0.679$, $\sigma_8 = 0.8344$, and $n = 0.9681$, which are taken from the most recent release of the \citet{2018arXiv180706209P}. Within a 5 Mpc$^3$ box with a root grid size of 128$^3$, a smaller 625 $\times$ 703.125 $\times$ 1093.75 kpc$^3$ sub grid encompassing the Lagrangian volume of a 10$^{10}\ \rm{M_{\odot}}$ halo (at $z=0$) is refined by a factor of 2$^4$ to create an effective grid size resolution of (2048)$^3$ with a dark matter particle mass of 1043 $\rm{M_\odot}$. Inside this smaller ``zoom-in" region, grids are allowed to further refine adaptively to up to a factor of 2$^{14}$ more than the root grid dimensions as successive density thresholds are exceeded. At $z=4$ this corresponds to a minimum proper cell width of $\sim 0.70$ pc, but typical values are between 11 and 180 proper parsecs within the virial radius of the largest halo at that redshift. The simulation includes 9-species (\hi, \hii, \hei, \heii, \heiii, e$^-$, H$_2$, H$_2^+$, H$^-$) radiatively-driven non-equilibrium chemistry, radiating star particles, and supernovae feedback \citep{2012MNRAS.427..311W} with the same parameters and thresholds described in \citet{2019MNRAS.tmp.2947B}.

To facilitate analysis of the time-dependence of emission line trends, the state of the simulation is saved every ~368,000 years starting at $z=6$ until $z=3$, which corresponds to about 3,400 outputs. In the simulation, a major merger ($\rm{M_\star} = 4.69 \times 10^{7}\ \rm{M_{\odot}}$ : $\rm{M_\star} =3.21 \times 10^{7}\ \rm{M_{\odot}}$) begins at $z \sim 4.17$ and concludes at $z \sim 3.5$. Therefore, two significant halos of roughly similar mass are available for study from $z =6$ until their merger. The larger halo at the time of merger and their resulting combined halo is henceforth referred to as Halo 0 and the smaller member of the merger will be referred to as Halo 1. At $z=6$, Halo 1 has almost twice the stellar mass as Halo 0 ($\rm{M_\star} = 1.63 \times 10^{7}\ \rm{M_{\odot}}$ versus  $\rm{M_\star} = 8.33 \times 10^{6}\ \rm{M_{\odot}}$), but subsequently exhibits a slower star formation rate. At $z = 3.5$, the stellar mass and total mass of Halo 0 grows to $2.01 \times 10^{8}\ \rm{M_{\odot}}$ and $1.23 \times 10^{9}\ \rm{M_{\odot}}$ respectively.

\subsection{Emission Line Model}

Emission lines are calculated in roughly the same manner as in \citet{2019MNRAS.tmp.2947B}, with some small improvements and additions. To summarize, a halo merger tree is produced by performing an iterative $r_{200}$ overdensity-finding algorithm tuned to return a consistent halo position and radius between timesteps as well as track Halo 1 through its merger. Then, using Flexible Stellar Population Synthesis \citep[FSPS;][]{2010ApJ...712..833C}, 8000-wavelength spectra are attached to each star particle based on its age, metallicity isochrome, and mass at each timestep.

From these spectra, a mean galactic spectrum is estimated and combined with the mean metallicity and density of the halo to produce wavelength-dependent absorption cross sections for the gas at 200 temperatures between 10$^{2.5}$ to 10$^{7}$ K using the {\sc Cloudy} photoionization solver \citep{2017RMxAA..53..385F}. These look up tables are separately generated for each halo and timestep and are attached to the combined mass of absorbers from simulation cells (\hi, \hei, \heii, H$_2$, H$_2^+$, H$^-$, and metals) as well as the cell temperature along rays to estimate the spectra and flux distribution within the halo. In a test, analytic models for the ionization cross sections of \hi, \hei, and \heii\ attached to the corresponding densities from the simulation for comparison. Therein, absorption along rays were roughly equivalent (within 5\%) to the {\sc Cloudy}-generated model at low to moderately high ionization fractions and a bit less absorptive at very high ionization fractions as the importance of \hi\ diminishes and other processes and species dominate the cross section. Because the cross section is only attached to the strong absorbers in the simulation and the cross sections are generated in the presence of the current galactic spectra, much of the non-equilibrium state of the simulation is preserved with this method, while additionally accounting for absorption phenomena that are not explicitly treated in the simulation.

Armed with a model for the attenuated spectra and flux at every point in a halo, a second round of {\sc Cloudy} calculations are used with a geometry prescription that matches the volume distribution of the flux within each cell to carefully account for the presence of multiple stellar sources as discussed in \citet{2019MNRAS.tmp.2947B}, and the resulting emission line luminosities are saved and reported. The prior study used a fixed photon path length to cell width ratio at this stage, whereas the cell photon path length used for the purposes of this calculation is estimated to be the luminosity-weighted mean path length from the stars through the cell to a point with low (1$\rm{^{st}}$\%ile) flux within the cell. Accordingly, the effective path length may be smaller than the smallest dimension of the cell up to the $\sqrt{3}$ times the width of the cell depending on the distribution of stars with respect to the cell as well as their luminosity.

\subsection{Escape Fraction and Dust Accounting}
\label{sec:escf}

\begin{figure*}
\begin{center}
\includegraphics[width=\linewidth]{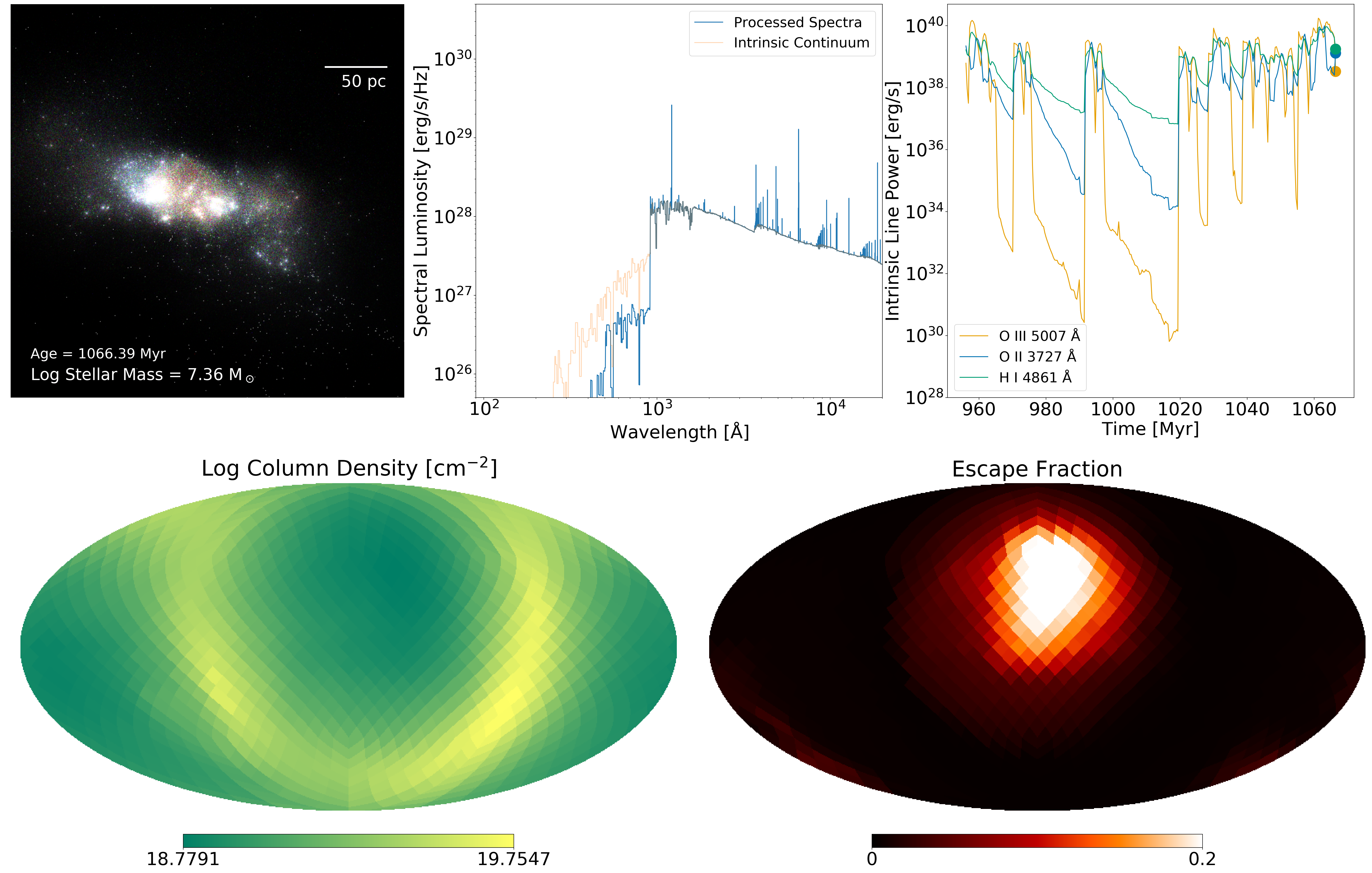}
\caption{Key galactic characteristics of Halo 0's evolution, demonstrating the analysis pipeline. From stellar populations, visualized as light sources in the top left dust and gas Monte Carlo ray tracing image, nebular emission lines are calculated throughout the halo to create a path-dependent processed spectra (seen in the middle top plot in a direction with high $f_{\rm esc}$). Emission line strengths are tabulated every 368 ky to determine the time-series trend shown in the top right plot. To determine continuum absorption, radiation is absorbed through the medium using a wavelength-dependent absorption profile that mostly depends on the neutral hydrogen column density (shown in the bottom left as a function of path to the virial sphere). The resulting ionizing escape fraction (shown in the bottom right) is reported as a distribution of 972 ray-traced paths from each star and reported as a luminosity weighted sum as further described in Sec. \ref{sec:escf}.}
%Key galactic characteristics during the only high $f_{\rm esc}$, low O$_{32}$ mock observation at 1066.39 Myr during Halo 0's evolution. Top row: left: true color composite dust Monte Carlo optical image showing one third of the virial diameter, middle: intrinsic stellar continuum (yellow) and attenuated spectra with emission lines (blue) at the virial radius in the direction of highest escape fraction, right: emission line history showing \oii\, \oiii\, and H$\beta$ line luminosities over the prior 100 Myr. Bottom row: anisotropic, polychromatic neutral hydrogen column density  (left) and ionizing escape fraction (right) generated by ray tracing from each source to the surface of the virial sphere and weighting the result in each of 972 directions by the luminosity of each source (see Sec. \ref{sec:escf}).}
\label{fig:snap2}
\end{center}
\end{figure*}

This work describes the relationship between O$_{32}$ and the escape fraction of ionizing radiation, $f_{\rm{esc}}$, which can be absorbed by both gas and galactic dust. The absorption cross sections used to attenuate stellar light for the emission line model include dust grain extinction, which is implicitly connected to the hydrogen nucleon column density through the use of the mean galactic metallicity in their determination. In this approximation, $f_{\rm{esc}}$ is calculated as

\begin{equation}
f_{\rm{esc}} = \sum_{k=1}^{\rm{N_{stars}}}\frac{\int_{\nu_l}^{\infty}L_{\nu,k}e^{-\tau_\nu}d\nu}{\int_{\nu_l}^{\infty}\sum_{j=1}^{\rm{N_{stars}}} L_{\nu,j}d\nu},
\end{equation}
\noindent
where $L_{\nu,k}$ is the spectral luminosity of star $k$ at frequency $\nu$ in units of erg s$^{-1}$ Hz$^{-1}$, $\nu_l$ is the frequency of the Lyman limit, and the optical depth, $\tau_\nu$, is defined as

\begin{equation}
\tau_{\nu} = \int_{\vec{r}_{\rm{star}}}^{\vec{r}_{\rm{rvir}}} \sigma(T(\vec{r}),\nu)\rho_{\rm{tot}}(\vec{r})d\vec{r},
\label{eq:tau}
\end{equation}
\noindent
where $\rho_{\rm{tot}}(\vec{r})$ is the summed density of strong absorbers at position $\vec{r}$ and $\sigma(T(\vec{r}),\nu)$ is the temperature-dependent mass attenuation coefficient at position $\vec{r}$ as well as frequency $\nu$. Eq. \ref{eq:tau} is a linear path integral from the position of each star particle, $\vec{r}_{\rm{star}}$, to a point on the surface of a sphere defined by the virial radius of the halo, $\vec{r}_{\rm{rvir}}$, along a vector drawn from one of 972 Hierarchical Equal Area isoLatitude Pixelations ({\sc HEALPix}) \citep{2005ApJ...622..759G} of a sphere. Thus, $f_{\rm{esc}}$ is a projection of onto the virial sphere in each {\sc HEALPix} direction, simulating parallel rays to an observer from each light source, but not accounting for emission and absorption outside of the virial radius. Depending on the direction, $f_{\rm{esc}}$ varies wildly owing to the non-homogeneous nature of optical depths among paths through simulated galaxies. While our calculation of escape fraction employs ray tracing at the best resolution available to the simulation, there is some evidence that further resolving cloud structures might affect the morphology of \hii\ regions immediately after star formation \citep[e.g.][]{2015MNRAS.454.4484G} and further investigations are needed to determine how this might affect galaxy escape fractions.

%as shown in the bottom right plot of Fig. \ref{fig:snap2}.

\begin{figure}
\begin{center}
\includegraphics[width=\linewidth]{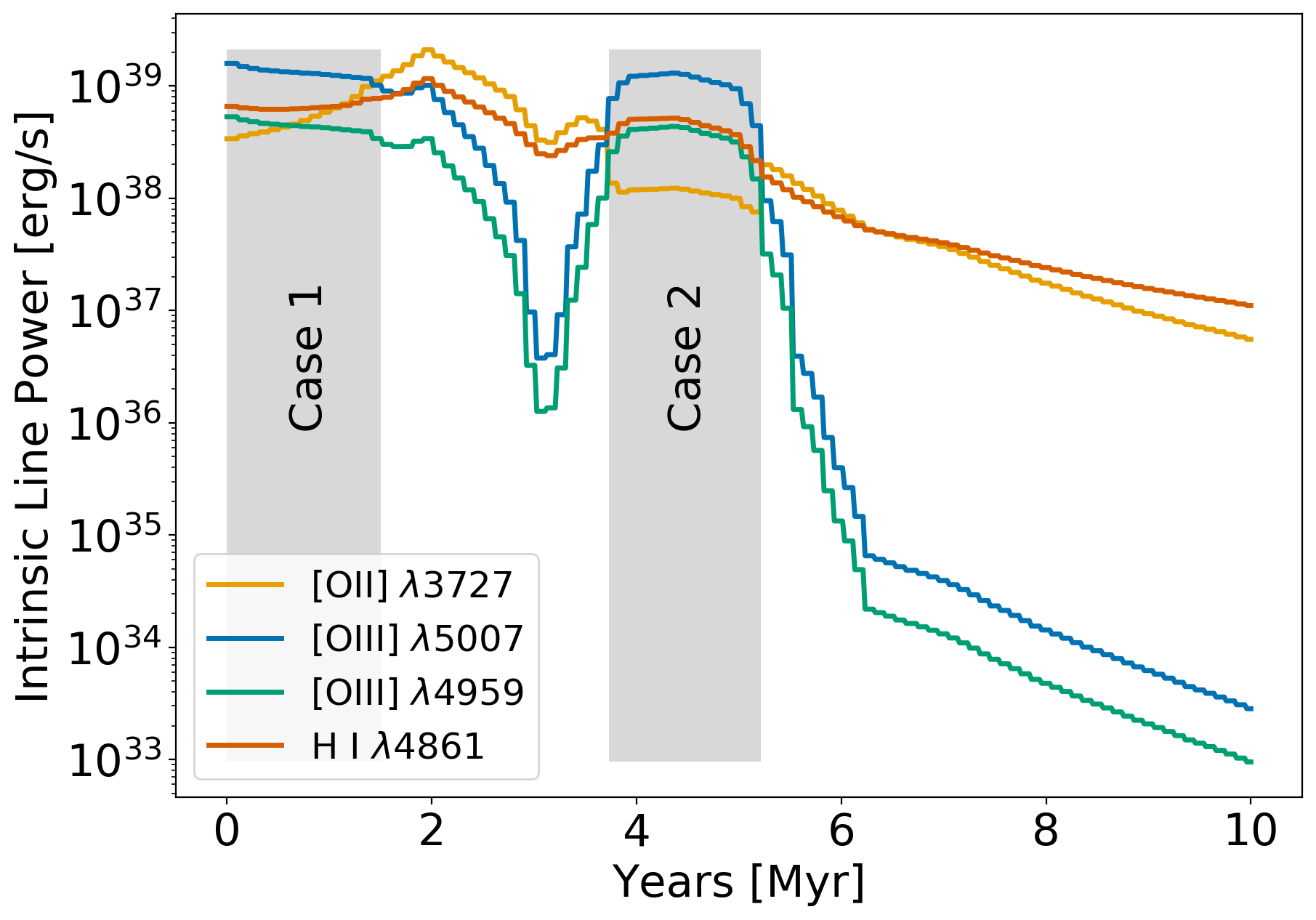}
\caption{Toy {\sc Cloudy} converged emission line pattern resulting from a single, isolated 10$^5\ \rm{M_{\odot}}$ starburst on a uniform sphere of gas with the same metallicity and density as Halo 0 at a simulation age of 1,096 Myr. Emission line powers are shown for \oiii, \oii, and H$\beta$ as the star cluster spectra evolves using FSPS and the star cluster mass evolves using the mass age relationship described in \citet{2005A&A...441..117L}. Regions where O$_{32}$ is greater than 1 after the onset of star formation (Case 1) and during the Wolf-Rayet phase (Case 2) are shaded in gray.}
\label{fig:indicluster}
\end{center}
\end{figure}

\begin{figure*}
\begin{center}
\includegraphics[width=\linewidth]{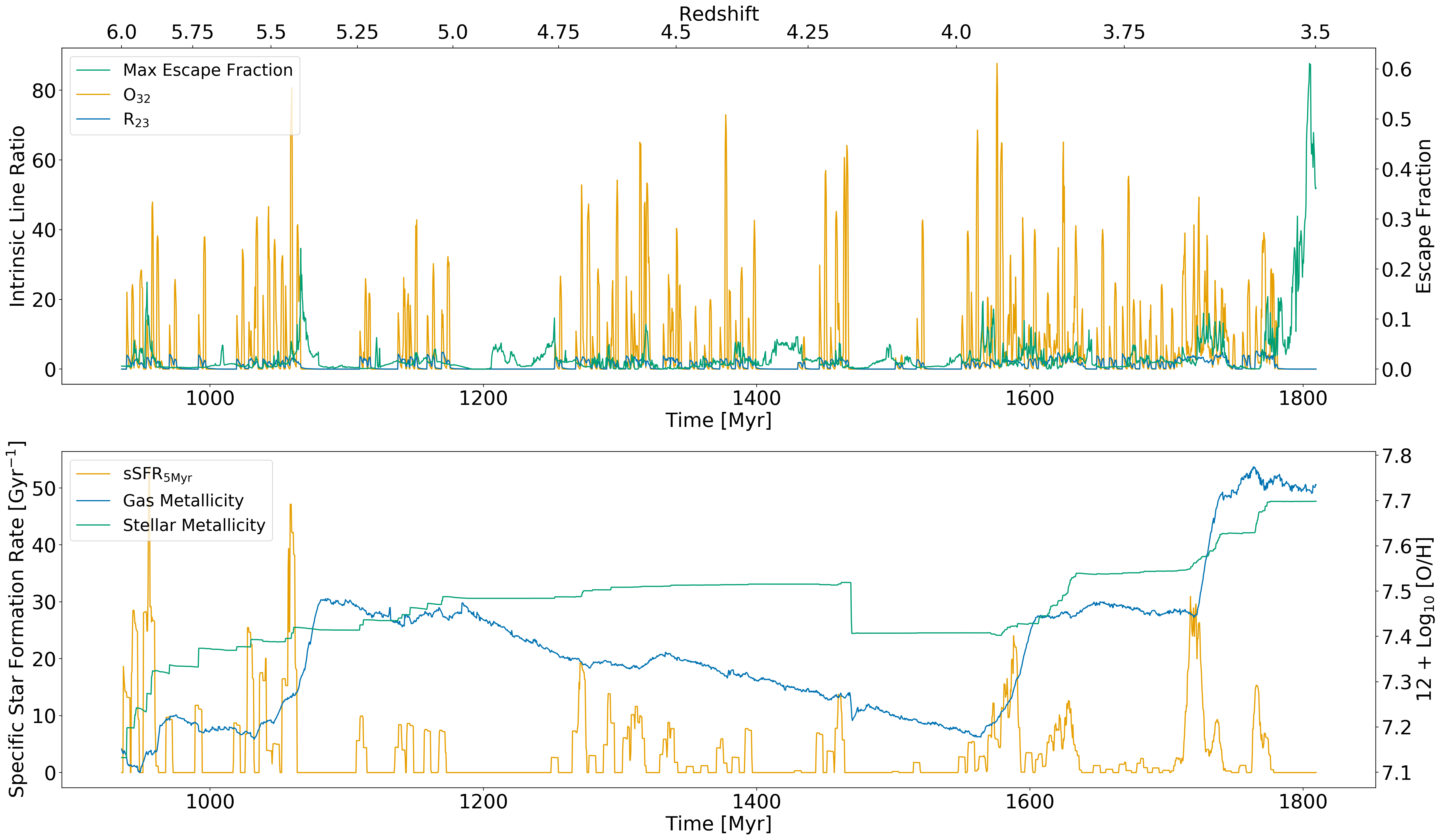}
\caption{Time series plot of the observables and galactic characteristics of Halo 0 plotted at a 386 kyr cadence. Top plot: The maximum escape fraction among 972 galaxy orientations (green), O$_{32}$ (salmon) and R$_{23}$ (blue) ratios. Bottom plot: 5 Myr-averaged specific star formation rate (yellow, left y-axis), stellar (green) and gas (blue) metallicity plotted on the right y-axis.}
\label{fig:fhalo0hist}
\end{center}
\end{figure*}

Because \oiii$\lambda$5007 and $\lambda$4959 are at longer wavelengths than the \oii$\lambda$3727 doublet and UV/optical dust extinction decreases as a function of wavelength, the presence of dust can increase O$_{32}$ at the virial radius relative to its intrinsic value in \hii\ regions. The boost in the ratio between lines at a wavelength A and a wavelength B due solely to dust extinction can be modeled as $B_{\rm{A/B}} = e^{\tau_{\rm{B}}-\tau_{\rm{A}}}$. If one assumes a similar path length from line emitting regions that produce line A and line B to the virial radius, in terms of the dust column mass density, $N_{\rm{dust}}$ in g cm$^{-2}$, and the mass attenuation coefficients, $\sigma$, the boost is $B_{\rm{A/B}} \leq (e^{N_{\rm{dust,max}}})^{\sigma_{\rm{B}}-\sigma_{\rm{A}}}$. Using the relationship between  neutral hydrogen nucleon column density, $N_{\rm{H}}$, and dust column mass density described in \citet{draine2011physics}, the luminosity-weighted maximum value of $N_{\rm{dust}}$ in either halo during the course of the simulation from sources to the virial radius among any of the {\sc HEALPix} directions is

\begin{equation}
 N_{\rm{dust},\rm{max}} \leq 0.0091\frac{[\rm{O/H}]_{\rm{max}}}{[\rm{O/H}]_\odot}m_{\rm{H}}N_{\rm{H},\rm{max}} \leq 4.4 \times 10^{-6}\ \rm{g\ cm ^{-2}},
\end{equation}
\noindent
where $m_{\rm{H}}$ is the mass of hydrogen and assuming a solar abundance of oxygen. Among the \citet{2003ARA&A..41..241D} $R_v$=3.1, 4.0 and 5.5 dust models, which span the gamut of dust grain compositions and sizes, the maximum value of $\sigma_{\rm{3727\AA}}-\sigma_{\rm{5007\AA}}$ is $\approx 5690\ \rm{cm}^{2}\ \rm{g}^{-1}$. This yields a maximum value of $B_{\rm{5007\AA/3727\AA}}$ of 1.025, or a 2.5\% percent boost. Because this calculation neglects dust scattering, which would further lower the boost by returning a fraction of scattered photons back to the line of sight, and also neglects evidence that lower metallicity galaxies like the halos at in this study have lower dust to $N_{\rm{H}}$ ratios \citep[e.g.][]{2014A&A...563A..31R,2018ApJ...855..133K},
the dust correction to the intrinsic O$_{32}$ is likely functionally negligible and certainly less than 2.5\%. Therefore, only intrinsic O$_{32}$ values are reported in this study. The same argument applies to R$_{23}$, since H$\beta$ is also optically thin and of intermediate wavelength between \oii$\lambda$3727 and \oiii$\lambda$4959.

Fig. \ref{fig:snap2} visually summarizes the data products from the pipeline during a key point in time where Halo 0 exhibits low O$_{32}$ and high ionizing continuum escape fraction, which is further explored and described in the results section. 

%\subsection{Radiative Transfer Post-Processing Model Summary}
%The top left image shows a true-color Monte Carlo post-processing image of the galaxy during an early-pre spiral phase showing dust, star-forming regions, and older populations of stars. The top middle plot shows the post-processed spectra, showing the direction with the highest escape fraction in blue over the intrinsic stellar continuum in yellow. The top right plot shows the time evolution of key emission lines (\oiii$\lambda$5007, \oii$\lambda$3727, and H$\beta$) for the halo during the prior 100 Myr. The bottom left plot demonstrates the anisotropic neutral hydrogen column density, which provides context for the anisotropic escape fraction shown on the bottom right. 

\subsection{Toy Cluster Model}

Since the emission lines derived from the simulation exist within a rapidly evolving cosmological environment, a toy model is also devised to clarify trends that exist within star clusters independently of galaxy dynamics. Using just an {\sc FSPS} model, {\sc Cloudy}, and a \citet{2005A&A...441..117L} cluster mass evolution prescription, trends in \oiii and \oii are computed and plotted in Fig. \ref{fig:indicluster} (see caption for more details).

At the onset of star formation, O$_{32}$ peaks above one since the \oiii\ emission peaks $\sim$1.5 Myr before the initial peak in \oii\ emission. In the range of ($\sim$3.7-5.2 Myr) after the cluster forms, a second, stronger O$_{32}$ peak occurs because the strength of the \oii$\lambda$3727 doublet decays over the first few million years after a star formation event and the harder spectra from the Wolf-Rayet phase of stars in the cluster suddenly converts the reservoir of \oii to \oiii. 

Since oxygen coincidentally has almost the same ionization energy as hydrogen, \oii$\lambda$3727 emission mirrors the evolution of the declining volume of the \hii\ region and thus closely matches the evolution and strength of H$\beta$ emission except during the Wolf-Rayet phase, where \oii$\lambda$3727 is further suppressed. These effects produce two classes of incidents of high O$_{32}$ ratios: Case 1 ($<$ 1.5 Myr) where \oii$\lambda$3727 doublet emission is strong and more likely to be detected, and Case 2 (3.7-5.2 Myr) where \oii$\lambda$3727 doublet emission is relatively weak and therefore less likely to be detected. In the interval between the cases, O$_{32}$ shortly falls to order unity before dipping further.

This toy model only calculates the contribution from a single instantaneous burst, but the broader simulation displays a tendency towards extended star formation events over tens of millions of years (see Fig. \ref{fig:fhalo0hist}, bottom plot showing specific star formation rates). Since several star particles are often formed in close spatial and temporal proximity in the simulation, each star formation event results in different emission line signatures due to the overlapping spectral phases of the contributing star particles during their evolution. The size and nature of their encompassing \hii\ regions may also play a role, which in turn may depend on prior star formation episodes. Therefore, the exercise of classifying  O$_{32}$ peaks in a galactic context is most appropriate in the case of isolated bursty star formation events or isolated star-forming regions and otherwise falls to degeneracies and stochastic peaks. It should also be noted that the existence of these cases is sensitive to the spectra assumptions of {\sc FSPS} and the earliest few million years after a cluster forms is a challenging modeling problem \citep[e.g.][]{2020arXiv200809780S}.

\section{Results}

Armed with a generalized radiative transfer model for the production of emission lines within a high-resolution, cosmological simulation of a observably large galaxy, the trends in O$_{32}$, R$_{23}$ escape fraction, and metallicity are described in the time domain to theoretically untangle observed correlations in high-redshift, high escape fraction galaxies.

\subsection{Time-Dependent Trends}

The topmost plot of Fig. \ref{fig:fhalo0hist} displays the evolution of Halo 0 with respect to maximum $f_{\rm{esc}}$, O$_{32}$, and R$_{23}$ and provides context for the time dependence of the relevant phenomena. Each star formation incident is proceeded by an initial short-lived peak in O$_{32}$ (Case 1), and followed by a subsequent stronger O$_{32}$ peak  (Case 2) as the spectra hardens during the Wolf-Rayet-powered phase of the star clusters. Bursts of star formation generate stellar and supernovae feedback that tempers subsequent star formation by photo-ionizing the ISM. Thus, high maximum $f_{\rm{esc}}$ values are rarely coincident with high O$_{32}$ as the former are tied to minima of the star formation burst cycle. This pattern is repeated in Halo 1 and is not dissimilar from $f_{\rm{esc}}$ patterns in \citet{2020MNRAS.tmp.2069M}.

%It should also be emphasized that these trends support the notion that a broad range of \oiii, \oii, and thus O$_{32}$ emission can be associated with an individual star formation event  %

% and so observations of O$_{32}$ are only tenuously connected to the underlying star formation rates in epochs of bursty star formation.

As shown the bottom plot of Fig. \ref{fig:fhalo0hist}, halo gas metallicity does not monotonically grow with stellar mass as low-metallicity gas inflows compete with enrichment from stellar feedback. The downward discontinuity in gas metallicity at $\sim$1470 Myr is an artifact of the redefinition of Halo 0 to include the merging, lower-metallicity Halo 1 and serves as a marker for the beginning of the merger process. During the merger, the galaxies make four relative periapsides with respect to each other before their bulges merge. Each periapsis drives a long, sustained episode of high star formation rates in both halos. This contrasts with the more sporadic bursts of star formation until the merger event and represents a distinctly dissimilar galactic environment to the pre-merger halos.

Gas mass fraction is physically connected to the escape fraction of ionizing radiation because it modulates the ionizing radiation density needed to create ionized channels through the halo. Low gas mass fractions are also correlated to low neutral hydrogen column densities (shown during the merger as a function of azimuthal and polar angle about the halo in the bottom left plot of Fig. \ref{fig:snap2}) and the peak and minimum column densities decrease by about one order of magnitude between $z=6$ and $z=3.5$. Halo 0's gas mass fraction within its virial radius declines from a high value of $f_{gas}=0.29$ at the beginning of the window down to $f_{gas}=0.16$ at the end of the window and then further declines to $f_{gas}=0.10$ at $z=3$. At $z=3.5$, $f_{\rm esc}$ reaches its highest values, though highly anisotropically, as high star formation rates feed ionizing radiation into a depleted reservoir of gas. In the 345 Myr interval between $z=3.5$ and $z=3.0$, no star formation occurs and the decrease the gas mass fraction is explained by the gas-poor accumulation of dark matter into the halo from the remnants of the merger environment.

\subsection{Relationship Between $f_{\rm esc}$ and O$_{32}$}

\label{fesco23}

%\begin{figure*}
%\begin{center}
%\includegraphics[width=\linewidth]{fouplot_0_1766.png}
%\caption{Caption}
%\label{fig:snap}
%\end{center}
%\end{figure*}

\begin{figure}
\begin{center}
\includegraphics[width=\columnwidth]{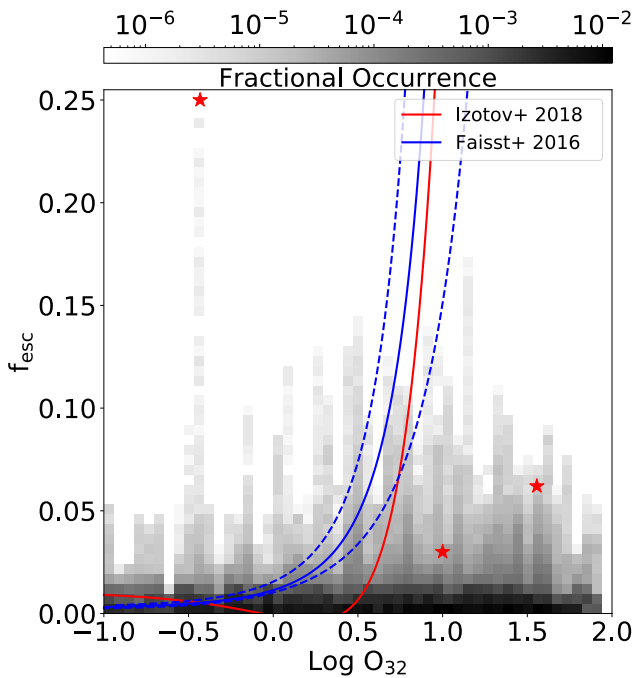}
\caption{Plot of possible mock observations of $f_{\rm esc}$ versus O$_{32}$ during the evolution of Halo 0 plotted as a gray-scale histogram. Also plotted are trends lines from \citet{2016ApJ...829...99F} (blue with lower and upper bounds in dashed lines) and \citet{10.1093/mnras/sty1378} (red). Outlying observations noted in Sec. \ref{fesco23} are shown as red stars.}
\label{fig:fesco32}
\end{center}
\end{figure}

\begin{figure}
\begin{center}
\includegraphics[width=\linewidth]{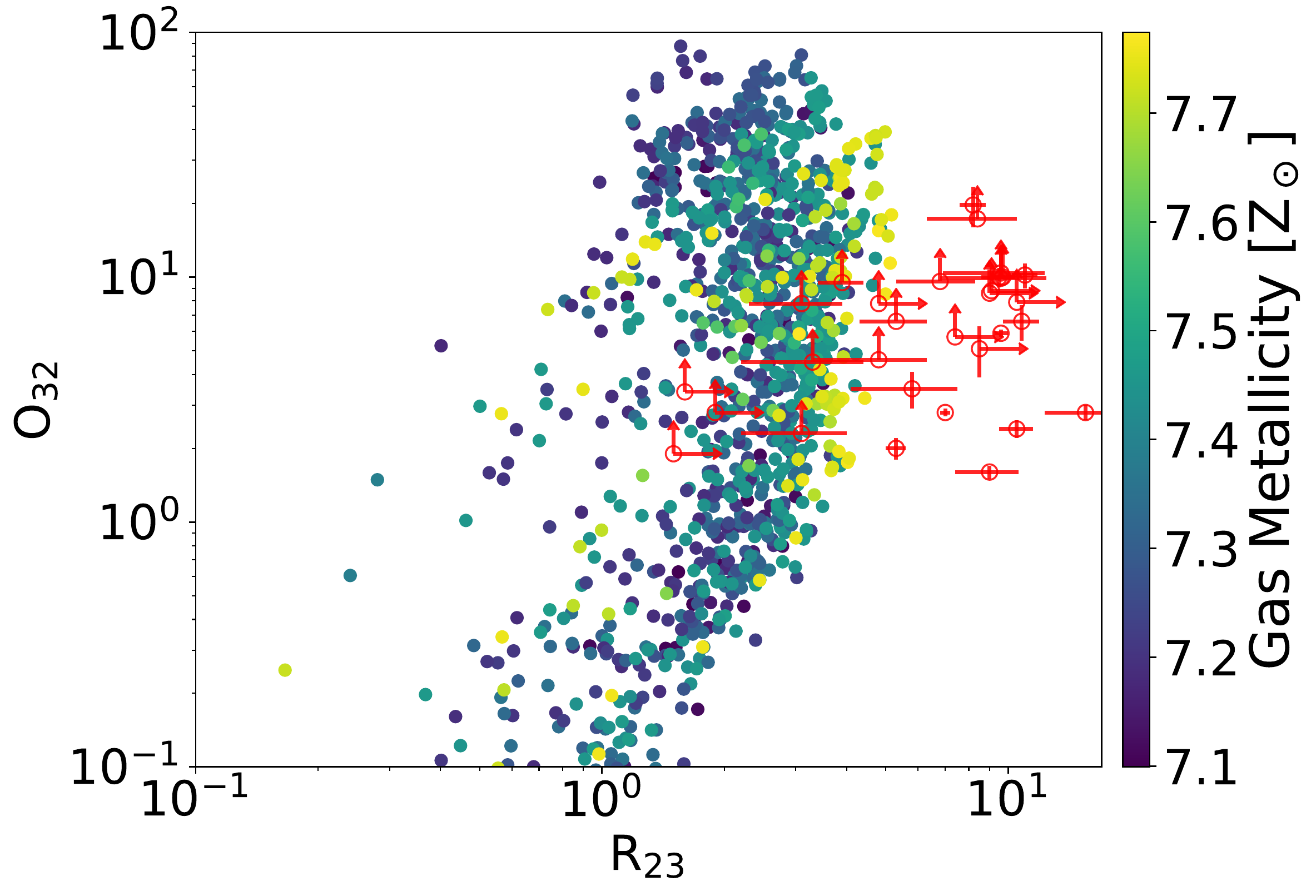}
\caption{R$_{23}$ versus O$_{32}$ trend of Halo 0 colored by metallicity (one point per timestep representing galaxy-wide line ratios) plotted with observations from \citet{nakajima2020a} (red). Halo 0's line ratios cover a wide parameter space, but lies to the left of some of the observations due to the relationship between R$_{23}$ and metallicity. Halo 1(not shown) has an even lower metallicity range (6.94 $\leq$ 12 + Log$_{10}$ [O/H] $\leq 7.24$) and its distribution is accordingly shifted to lower R$_{23}$ by $\sim0.2$ dex, while maintaining roughly the same overall shape.}
\label{fig:o23r23}
\end{center}
\end{figure}

In Fig. \ref{fig:fhalo0hist}, there is a clear offset between the short period bursts of O$_{32}$ and the longer period peaks of $f_{\rm esc}$, however a positive correlation between observed $f_{\rm esc}$ and O$_{32}$ has been described in the literature. In this section, observations in the literature are detailed and then compared to synthetic observations calculated from the simulation to determine whether observed trends can be explained by time-dependent phenomena.

Fig. \ref{fig:fesco32} shows observationally-inferred trends between $f_{\rm esc}$ and 
$O_{32}$ (blue and red lines) \citep{2016ApJ...829...99F,10.1093/mnras/sty1378}. In addition to galaxies that fall within these trends, there exists examples of galaxies with low $f_{\rm esc}$ and high O$_{32}$ such as J1011+1947, which has an O$_{32}\sim 36$ and $f_{\rm esc}=0.062$ or $f_{\rm esc}=0.114$ depending on the estimation method, or J1248+4259, which has an O$_{32}>10$ and
$f_{\rm esc}\lesssim0.03$ \citep{10.1093/mnras/sty1378}. Recently, \citet{bassett2019a} added a single example (ID: 17251) of low O$_{32}$ (0.37) and high $f_{\rm esc}$(0.25) at $z\approx 3$ to the literature and challenged the notion that there is a trend between the two variables at all. These off-trend observations are indicated as red stars in Fig. \ref{fig:fesco32}.
%There are no observed examples of galaxies with low O$_{32}$ ($<$2) and high $f_{\rm esc}$ ($>0.1$). 

%The association between the Lyman continuum escape fractions computed at each time in our study (the top panel of Fig. \ref{fig:fhalo0hist} plots the simulated trends in the maximum escape fraction for Halo 0 and the minimum value is always close to zero) and the corresponding line ratios are further explored to compare with the observed trends. 

In the simulations of Halo 0 and Halo 1, observations of $f_{\rm esc}$ are both time-dependent and galaxy orientation-dependent as high $f_{\rm esc}$ values only escape into the IGM in highly focused channels (as shown in the bottom left plot of Fig. \ref{fig:snap2}), resulting in a distribution of possible observations at each time step. As described in Section \ref{sec:escf},
to compute this distribution,
O$_{32}$ values at each timestep are associated with 972 corresponding $f_{\rm esc}$ values in each of the {\sc HEALPix} directions, resulting in more than two million possible combinations for Halo 0 between $z=6$ and $z=3.5$.
Combinations of mock observations of O$_{32}$ and $f_{\rm esc}$ are then histogrammed by fractional occurrence and shown in gray scale in Fig. \ref{fig:fesco32}. 

\subsubsection{Outlying Combinations of O$_{32}$ and $f_{\rm esc}$}

The resulting simulated distribution roughly traces the distribution of observations including an outlying set of low O$_{32}$ ($<$ 2) and high $f_{\rm esc}$ ($>0.1$) values similar to the outlying observation of 17251 from \citet{bassett2019a}. Here we explore why this is rarely observed and falls outside the distribution of the rest of both the simulated and observed data.
%Halo 0's evolution roughly traces the distribution of observations except for the absence of examples of very high $f_{\rm esc}$ values ($>$0.15). However, this only considers galaxies with M$_\star$ $>$ 10$^{7.6}$ M$_\odot$, which is less than the minimum calculated mass in the observed sample.

Halo 0's examples of low O$_{32}$ (0.34) and high $f_{\rm esc}$ ($\leq$ 0.24) occur at a single timestep of the 3,400 studied (t = 1066.39 Myr, M$_\star$ = 10$^{7.36}$ M$_\odot$), which can be seen in the top plot of Fig. \ref{fig:fhalo0hist},  Fig. \ref{fig:snap2}, and Fig. \ref{fig:fesco32}.
This instance is about 5.5 Myr after the 5 Myr-averaged specific star formation rate of the halo reaches its highest value over the interval at 42.22 Gyr$^{-1}$ (30.51 Gyr$^{-1}$ when averaged over 10 Myr), which corresponds to the end of the Case 2 phase of the O$_{32}$ emission pattern.
At time $t=1066.39$ Myr, \oiii\ and \oii\ emission line luminosities are falling rapidly and O$_{32}$ is itself falling at a rate of about one order of magnitude every 368 kyr timestep. This decline follows several consecutive bursts of rapid star formation, when the distribution of gas in the galaxy is morphologically irregular (as seen in the true-color photon Monte Carlo image in the top left of Fig. \ref{fig:snap2}).
Gaps in the gas open in the wake of strong supernovae feedback from prior star formation events, and enable both
an abnormally large \hii\ region and a wide, ionized channel to the virial radius.

The highest escape fraction ($f_{\rm esc}=0.077$) in Halo 1 registered when O$_{32}$ $< $ 2,
but this occurs shortly before the first infall of the merger at $z=4.27$ and may not be
independent of the interaction with Halo 0.
That incident is, however, also preceded by a $\sim$30 Myr period of sustained star formation with peaks in 5 Myr-averaged specific star formation rates in excess of 15 Gyr$^{-1}$ that disrupt and precondition the gas for the formation of a larger \hii\ region.

%If one disregards the extremely and perhaps undetectably rapid rise in O$_{32}$ at the birth of cluster, 
%There are two instances of low but observable O$_{32}$ in these data that could, in principle, be associated with high $f_{\rm esc}$ after a star formation event. The first is during the drop in O$_{32}$ between the Case 1 and the Case 2 phase, and the second is at the end of the Case 2 phase. As shown in Fig. \ref{fig:indicluster} and Fig. \ref{fig:fhalo0hist}, both of these possibilities occur with a shorter delay after cluster formation than the subsequent rise in $f_{\rm esc}$ and so the combination in question only occurs after a long period of sustained star formation preconditions the gas, but before feedback shuts down local star formation.

With sustained star formation, the minimum in O$_{32}$ between the Case 1 and Case 2 phases is boosted by the constructive sum of stars in various phases of their evolution. Though most star formation events produce multiple star particles in our simulation, real clusters would likely have an even wider range of stellar ages that would further boost the O$_{32}$ minima between Case 1 and 2.
Therefore, the drop in O$_{32}$ at end of the last Case 2 phase, which occurs at the end of a period of sustained star formation, presents the best opportunity for high $f_{\rm esc}$ with low O$_{32}$ to be observed. However, that combination of conditions is several times rarer and more transient than other cases where O$_{32}$ and/or $f_{\rm esc}$ can be observed. In the case of Halo 0, the combination of a large specific star formation rate after a long period of sustained star formation in an irregular and small galaxy suggests that the necessary conditions to produce low O$_{32}$ and high $f_{\rm esc}$ were not impossibly rare though.% and the resulting emission line powers were weak and in a regime that would be difficult to detect.

%as well as occurs when the M$_\star$ $< $10$^{7.6}$ M$_\odot$.

From $z=6$ until $z=3$, less than one in one hundred thousand synthetic observations of Halo 0 or Halo 1 had a low (0.01-2) O$_{32}$ and high ($>$0.1) $f_{\rm esc}$. However, more than one in three hundred combinations of high ($>$0.1) $f_{\rm esc}$ and low to nonexistent ($<$0.01) O$_{32}$ occurred. These fractions do not correspond to observational probabilities since they do not take into account telescope sensitivity limits and come from a limited sample of galaxies, but do elucidate trends that provide context to the current array of observations. Taken together, the top left region of Fig. \ref{fig:fesco32} (low O$_{32}$ and high  $f_{\rm esc}$) is almost completely depopulated for three reasons: 1) the phase offset between high $f_{\rm esc}$ and the presence of O$_{32}$ due to feedback cycles, 2) the beaming of $f_{\rm esc}$ through ionized channels reducing the overall probability of high $f_{\rm esc}$ observations, and 3) the rarity of cases of low O$_{32}$ due to it being a more transient state than non-existent or high O$_{32}$. Conversely, in rare exceptions, outlier values in this region can be produced when conditions align. It should be noted that neither halo explores the full parameter space of observations or conditions as evidenced by the absence of incidences of higher $f_{\rm esc}$ like those reported in \citep{nakajima2020a} in Fig. \ref{fig:fesco32}. A larger sample of simulated galaxies would be needed to make more general inferences about the observed rates.

%This study shows, however, that maximum observable value of escaping ionizing intensity from Halo 0 scales poorly with observed escape fractions and underlines the care needed to infer intrinsic escape fractions from continuum emission observations.

Importantly, since O$_{32}$ emission is optically thin and $f_{\rm esc}$ is anisotropic, there is $no$ $direct$ $causal$ $connection$ between O$_{32}$ and $f_{\rm esc}$ in the simulations, despite the trend in observations.
Each incident of O$_{32}$ emission corresponds to a range of possible $f_{\rm esc}$ depending on the observer's orientation with respect to the galaxy and $f_{\rm esc}$ peaks occur on a longer timescale after star formation than an O$_{32}$ peak.

%However, high observable $f_{\rm esc}$ occurs when O$_{32}$ is also high. Indeed, it is likely for 

\subsection{Relationship Between O$_{32}$, R$_{23}$, and Metallicity}

Under the assumption of solar abundances, Halo 0 non-monotonically traverses a 12 + Log$_{10}$ [O/H] gas metallicity range from 7.10 to 7.77 over the stellar mass range of 6.92 $<$ Log [M$_\star$/M$_\odot$] $<$ 8.31 as it evolves from $z=6$ to $z=3.5$. Compared to the sample of lower mass local galaxies from the \citet{2013ApJ...765..140A} SDSS metallicity-mass relationship, metallicities in this much higher redshift simulation scatter above and below the mean of the observed trend with a bias towards lower metallicities, especially at higher stellar masses.

The line ratio R$_{23}$  is often used to estimate the metallicity of observed galaxies.
% In the observed sample of low redshift galaxies ($z <0.4$), the ratio peaks at mean of around R$_{23}\sim12.5$ at 12 + Log$_{10}$ [O/H] $\sim 8.0$  \citep{2006A&A...459...85N,2019ApJ...872..145J}. In the same studies, R$_{23}$ in a sample of extreme emission line galaxies declines to range between R$_{23}\sim4.5-6$  for 12 + Log$_{10}$ [O/H] $\sim 7.4$, with significant scatter given the small sample size at those metallicities.
Recently, data collected for high redshift ($z >3$) high EW \oiii\ sources showed they occupy a range of R$_{23}\sim 1.5-15.5$ \citep{nakajima2020a}, as plotted with error bars in red on Fig. \ref{fig:o23r23}. Some of the lowest R$_{23}$ ratios are lower bounds due to H$\beta$ and their true R$_{23}$ may be larger. Similarly, the lower bounds in O$_{32}$ plotted in Fig. \ref{fig:o23r23} could be significantly less than the true value owing to the presence of \oii$\lambda$3727 in the denominator.
%Some of the lowest R$_{23}$ ratios are lower bounds due to \oii$\lambda$3727 non-detections, and their true R$_{23}$ may lie upward by as much as a factor of two. Similarly, the lower bounds in O$_{32}$ plotted in Fig. \ref{fig:o23r23} could be significantly less than the true value owing to the presence of \oii$\lambda$3727 in the denominator.

To compare with the observations, synthetic R$_{23}$ and O$_{32}$ data from Halo 0 are
shown in Fig. \ref{fig:o23r23}, and colored by metallicity.
The simulated galaxies at $z\sim4$ have lower metallicities than many of the systems in the $z\sim3$ \citet{nakajima2020a} sample, and both Halo 0 and Halo 1 (not shown) are offset to lower R$_{23}$
given their lower metallicities. The distribution of O$_{32}$ vs. R$_{23}$ values during the evolution
of the simulated galaxies is more complicated, and properties beyond metallicity influence its
behavior. While the highest O$_{32}$ values in Halo 0 occurred at low metallicities and low R$_{23}$, Halo 1 displayed coincident peaks in O$_{32}$ and R$_{23}$ during its evolution.
This difference suggests that where the maximum of O$_{32}$ occurs relative to R$_{23}$
is also connected to, e.g., a galaxy's specific star formation rate in addition to metallicity.
For instance, during individual starbursts when there is a highly variable specific star formation
rate, the O$_{32}$ and R$_{23}$ line ratios can rapidly move between low O$_{32}$-low R$_{23}$ and
high O$_{32}$-high R$_{23}$ states, and can even show R$_{23}<1$ for short intervals. However, scatter in O$_{32}$ gradually decreases with increasing gas metallicity.

In summary, while our simulation reproduces many of the LACES observations of R$_{23}$ and O$_{32}$, R$_{23}$ analysis suggests that a higher gas metallicity simulated sample would be needed to cover the full range. Additionally, there is only weak evidence from our simulation that O$_{32}$ and R$_{23}$ or O$_{32}$ and gas metallicity are clearly correlated in this metallicity regime and significant and time-dependent and metallicity-dependent scatter in those relationships exists due to other processes. However, as also seen observations \citep[e.g.][]{2019A&ARv..27....3M}, maximum values of O$_{32}$ decrease with increasing metallicity and R$_{23}$. While that relationship might help guide comparisons between the work and higher metallicity observations, the scatter inherent in our synthetic line ratio calculations makes those comparisons challenging. 

\section{Discussion and Conclusions}

Using high-resolution zoom-in simulations of star-forming galaxies at $z\sim4$, a radiative transfer
post-processing is used to explore the time evolution of the emission line ratios
O$_{32}$ and R$_{23}$, and the Lyman-continuum escape fraction $f_{\rm esc}$.
In summary, our key findings are:
\medskip

\begin{itemize}

\item The simulations predict that high escape fraction (e.g., $f_{\rm esc}>0.05$) is almost always accompanied by high oxygen emission line ratios (e.g., O$_{32}>3$). However,
while $f_{\rm esc}$ and O$_{32}$ are both powered by a hard ionizing spectra, the response time for the creation of an ionized channel that allows for high $f_{\rm esc}$ is much longer than the production of a high value of O$_{32}$, and thus the two phenomena are not causally related.

\item The combination of a low value of O$_{32}$ and a high $f_{\rm esc}$ is likely a rare event that occurs at the end of a long burst of star formation and persists for only a few hundred thousand years.

\item Metallicity is degenerate on an O$_{32}$ versus R$_{23}$ plot due to tendency for the galaxy to move diagonally through the plane during star formation events.

\end{itemize}

Though this study was also able to explore more of the galaxy-scale dynamical nebular emission line parameter space than prior studies, our sample galaxies only occupy a portion of the observational space.
A complementary
recent study explored the statistics of these quantities in static outputs of multi-galaxy simulations \citep{katz2020a}, and the results are broadly consistent despite this studies focus on examining a the time-evolution of only a pair galaxies. Given the importance of the
relative time evolution of $f_{\rm esc}$ and O$_{32}$, a larger sample of simulated galaxies with sufficient time cadence to make statistical arguments about the nature and evolution of nebular emission lines is still required and will be explored in future work.

\section*{ACKNOWLEDGMENTS}

This work was supported by XSEDE computing grants TG-AST190001 and TG-AST180052 and the Stampede2 supercomputer at the Texas Advanced Computing Center. KSSB was supported by a Porat Postdoctoral Fellowship at Stanford University.
BER was supported in part by NASA program HST-GO-14747, contract NNG16PJ25C, and grant 80NSSC18K0563, and NSF award 1828315. RSE and AS acknowledge funding from the European Research Council under the European Union Horizon 2020 research and innovation programme (grant agreement No 669253). We acknowledge use of the lux supercomputer at UC Santa Cruz, funded by NSF MRI grant AST 1828315. 

\bibliography{combined}{}
\bibliographystyle{aasjournal}

\end{document}